\documentclass[12pt]{iopart}
\usepackage{enumitem}
\usepackage{graphicx}
\usepackage[utf8]{inputenc}
\usepackage[svgnames]{xcolor} 
\usepackage{lipsum}
\usepackage[super,sort,compress]{cite}
\usepackage{iopams} 
\expandafter\let\csname equation*\endcsname\relax
\expandafter\let\csname endequation*\endcsname\relax
\usepackage{amsmath}
\usepackage{bm}
\usepackage{hyperref}
\usepackage[utf8]{inputenc}
\usepackage[svgnames]{xcolor} 
\usepackage{lipsum}

\begin{document}
\setlength{\abovedisplayskip}{3.5pt}
\setlength{\belowdisplayskip}{3.5pt}

\title{Joint quasiprobability distribution on the measurement outcomes of MUB-driven operators}


\author{H S Smitha Rao$^1$, Swarnamala Sirsi$^1$ , 
Karthik Bharath$^2$}

\address{$^1$ Yuvaraja's college, University of Mysore, Mysuru, India }
\address{$^2$ University of Nottingham, Nottingham, U.K.}
\ead{smitharao@ycm.uni-mysore.ac.in}

\section*{Abstract} 
We propose a method to define quasiprobability distributions for general spin-$j$ systems of dimension $n=2j+1$, where $n$ is a prime or power of prime. The method is based on a complete set of orthonormal commuting operators related to Mutually Unbiased Bases which enable (i) a parameterisation of the density matrix and (ii) construction of measurement operators that can be physically realised. As a result we geometrically characterise the set of states for which the quasiprobability distribution is non-negative, and can be viewed as a joint distribution of classical random variables assuming values in a finite set of outcomes. The set is an $(n^2-1)$-dimensional convex polytope with $n+1$ vertices as the only pure states, $n^{n+1}$ number of higher dimensional faces, and $n^3(n+1)/2$ edges.

\section{Introduction} 
Expectation values of quantum mechanical observables can be studied on the continuous phase space using quasiprobability distributions (QPDs). For spin systems, prominent ones include the Wigner \cite{Hillery} and Margenau-Hill \cite{MHill} QPDs; in quantum optics, the Glauber and Sudarshan\cite{ECG} QPD for quantum radiation is used. These behave like probability density functions on the continuous phase space, but, by their very definition, can assume negative values in certain regions of the phase space. The issue also afflicts versions of the Wigner QPD on discretised phase space\cite{wignerwoot, gibbons04}, tailored for finite-dimensional quantum systems. Occurrence of negative values of QPDs, that can be attributed to non-commutativity of quantum mechanical observables, is used as a signature of non-classicality \cite{class06,Spekkens08,ryu2013,ryu2019} and is exploited profitably in various quantum computational tasks\cite{Galvao04,Veitch12,rebit,rauss17}.

For finite-dimensional quantum systems, QPDs unrelated to the phase space view have also been proposed \cite{Devi1994, ryu2013,ryu2019} with a view towards ascribing joint probabilities to a finite set of measurement outcomes using measurement operators. The choice of measurement operators then plays an important role in examining properties of the corresponding joint QPD: the form of the QPD can be used to identify the subset of states for which the QPD behaves, and can be interpreted, as a classical joint distribution. It is desirable to consider measurement operators that are related to a complete set of Mutually Unbiased Bases (MUBs)\cite{woot}; this is the approach adopted while defining the discrete Wigner QPD \cite{wignerwoot, gibbons04,Galvao04}. For qubit or spin-1/2 systems, the Pauli operators can be used
since their eigenbases constiute a complete set of MUBs. In higher prime or power-of-prime dimensions, the eigenbases of the  generalised Pauli operators\cite{Bando} can be chosen as the MUB basis vectors, but the operators themselves are not observables. 

The two-fold purpose of this paper is: (i) to construct QPDs for $n$-dimensional quantum systems, or equivalently spin-$j$ systems with $n=2j+1$, where $n$ is a prime or power of a prime, using $n+1$ measurement operators $M_1,\ldots,M_{n+1}$ related to a complete set of MUBs; (ii) obtain a geometric description of the set of states for which the constructed QPD corresponds to a valid joint distribution on measurement outcomes. The two objectives are achieved by employing an orthonormal operator basis given by the complete set of commuting operators (CSCOs) $\mathcal A=\{\hat \alpha_1,\ldots,\hat \alpha_{n^2-1}\}$ proposed in our earlier work\cite{SRao19}; the operators, through their relation to MUBs, can be partitioned into $n+1$ disjoint subsets $\mathcal A_1,\ldots,\mathcal A_{n+1}$, such that each $\mathcal A_i$ contains exactly $n-1$ operators that commute and permit simultaneous measurements. This enables us to construct  the measurements $M_i$ as linear combinations of the commuting operators in $\mathcal A_i$. A striking feature of the constructed QPD is that when restricted to each $\mathcal A_i$ it is a valid joint distribution of corresponding measurement outcomes for \emph{any} state. 

Our method of construction of the QPD uses a characteristic function defined for a density matrix using a particular operator ordering of the CSCOs. As a consequence, for states for which the QPD is non-negative it can effectively be viewed as a joint distribution of classical random variables $X_1,\ldots,X_{n^2-1}$ assuming values on a finite set of outcomes. Relatedly, bivariate and trivariate probability distributions for outcomes from Pauli measurements corresponding to different definitions of characteristic functions have been derived for qubits \cite{Chandler92, Cohen1986}, and characteristic functions have also been used to derive trivariate moments for arbitrary spin-$j$ systems\cite{GR,trivariate97}. 

The use of CSCOs is well-motivated: unlike general $\mathfrak{su}(n)$ generators (e.g. Gell-Mann matrices) they are better suited for physical implementation and interpretation owing to their relationship with MUBs; 
they enable us to uncover the geometry of the set of states for which the QPD is non-negative. The set of such states forms a regular convex polytope on $n(n+1)$ vertices, $n^{n+1}$ number of higher dimensional faces, and $n^3(n+1)/2$ edges with each vertex on the surface of the Bloch ball of radius $\sqrt{n-1}$ which represents the set of all states. Interestingly, our method of using MUBs results in the convex polytope which coincides with the polytope identified by Galv\~{a}o \cite{Galvao04,class06} using probability coordinates obtained from definition of the discrete Wigner QPD under a phase space description. We briefly comment on this relationship (Section \ref{sec:spin1geom}), and leave detailed investigations for future work.  

We first review quantum characteristic functions (Section \ref{sec:qcf}). Then, starting with a brief description of how the program is carried out for spin-1/2 systems (Section \ref{sec:spin1/2}), we provide a detailed description of the QPD construction, corresponding geometry and physical realisation for spin-1 systems (Sections \ref{sec:csco} and \ref{sec:spin1}), and provide some details for spin-3/2 systems (Section \ref{sec:spin3/2}). Inspection of these two cases will reveal how the methodology extends to arbitrary spin-$j$ systems (Section \ref{sec:spinj}). 
 

\section{Quantum characteristic functions}
\label{sec:qcf}
For a $k$-dimensional classical random vector $\vec Y$ with joint distribution $p(\vec y)=p(y_1,\ldots,y_k)$, the Fourier transform 
\[\phi(\vec t)=\int e^{i\vec{t}\cdot\vec{y}}p(\vec {y})\text{d} \vec y \quad \text{or} \quad
\phi(\vec t)=\sum_{\vec y}e^{i\vec{t}\cdot\vec{y}}p(\vec {y})
\]
is referred to as its characteristic function, depending on whether $p(\vec y)$ is a continuous or discrete distribution. A characteristic function uniquely determines $p$ through its inverse Fourier transform. If we view a vector  $\vec X=(X_1,\ldots,X_k)$ of measurement operators $X_k$ as a quantum analogue of a classical random vector, noncommutativity implies that there are multiple ways to  define $e^{\vec t\cdot\vec X}$, and hence the characteristic function $\phi$ \cite{cohen2020}. This problem is typically addressed using symmetrisation rules, popular amongst which are the Margenau-Hill \cite{MHill} rule, which, for example when $k=3$, proposes
\begin{align*}
\label{MM}
e^{i\vec t\cdot\vec X} \longrightarrow
 \frac{1}{3!} \sum_{\pi \in \Pi_3} \Big[e^{it_{\pi(1)}{X_{\pi(1)}}}
 e^{it_{\pi(2)}{X_{\pi(2)}}}e^{it_{\pi(3)}{X_{\pi(3)}}}\Big],
\end{align*}
where $\Pi_k$ is the symmetric group of permutations of $[k]=\{1,\ldots,k\}$ with bijections $\pi: [k] \to [k]$, and the Wigner-Weyl \cite{wigne} rule, which proposes $e^{i\vec t\cdot\vec X} \to e^{i\vec t\cdot\vec X}$ for any fixed chosen ordering of $X_1,\ldots,X_k$. Note that $\vec X$ need not be a POVM for such a definition of $\phi$. 
For a chosen symmetrisation rule the quantum characteristic function associated with a state $\rho$ and operators $\vec X$ is then defined as $$\phi(\vec t)=\Tr [\rho e^{i\vec t\cdot\vec X} ].$$

Irrespective of the symmetrisation rule, the map $ \vec t \mapsto \phi(\vec t)$, unlike the situation with classical random variables, is not guaranteed to be the Fourier transform of a joint probability distribution $p(\vec x)$ on measurement outcomes for every state $\rho$\footnote{This is consequence of Bochner's theorem: $\phi(\vec t)$ is a valid characteristic function if and only if for every $r$-tuple $(\vec t_1,\ldots, \vec t_r)$ the $r \times r$ matrix with entries $\phi(\vec t_i -\vec t_j), i,j=1,\dots,r$ is non-negative definite and Hermitian. See Example 4.1 in \cite{partha} for a detailed discussion of the issue.}. However, $\phi$ can be inverted to obtain a QPD on the measurement outcomes. We will use the QPD arising from the Margenau-Hill symmetrisation rule using measurement operators constructed using the CSCOs. 
 \section{Spin-1/2 system}
 \label{sec:spin1/2}
It is instructive to first describe our construction for the spin-1/2 case with Pauli operators. The density matrix assumes the form
$\rho(\vec \theta)=\frac{1}{2}(\mathbb{I}_{2}+ \vec{\sigma} \cdot \vec \theta)$ where $\vec \sigma=(\sigma_x,\sigma_y,\sigma_z)$ with $\sigma_{i}$, $i=1, 2, 3$ denoting the  well-known Pauli operators, and the components of Bloch vector $\vec{\theta}$ are such that $\theta_{i}= Tr(\rho (\vec \theta)\sigma_i)$. The constraint $Tr[\rho(\vec \theta)^{2}] \leq 1$ implies that $\theta_{1}^{2} + \theta_{2}^{2} + \theta_{3}^{2}  \leq 1 $, with equality attained only for pure states. The set of density matrices for qubits is then the Bloch ball $\mathcal B^2(\vec \theta):=\{\vec \theta: \theta_{1}^{2} + \theta_{2}^{2} + \theta_{3}^{2}  \leq 1 \}$ with the surface of the sphere corresponding to pure states. 

It is known that the optimum measurement operators based on spin-1/2 MUBs can be constructed and physically realized using the Stern-Gerlach experimental setup. In this case, a particle having magnetic moment $\vec{\mu}$ is passed through an inhomogeneous magnetic field $\vec{B}$. Here the potential energy associated with the particle is $\mathcal{H}= -\vec{\mu}. \vec{B}$, where $\vec{\mu}$ is proportional to spin. When the magnetic field is oriented along the z-direction, one can measure the expectation value of $\sigma_3$. The corresponding Hamiltonian is $\mathcal{H}=h_0\mathbb{I}+h_3 \sigma_3$, whose expectation value results in $\langle \mathcal{H} \rangle= \Tr(\rho \mathcal{H})= h_0+\theta_3 h_3$. Then, the observable $\sigma_1$ can also be measured using the same apparatus if its diagonal basis has the same form as $\sigma_3$. Experimentally this corresponds to the application of magnetic field along x-direction. Similarly is the measurement of $\sigma_2$. This results in the complete determination of parameters characterizing the spin-1/2 density matrix. 

\subsection{Quasiprobability distribution and geometric description of non-negative region}
\begin{figure}[!ht]
\centering
\includegraphics[width=0.35\textwidth]{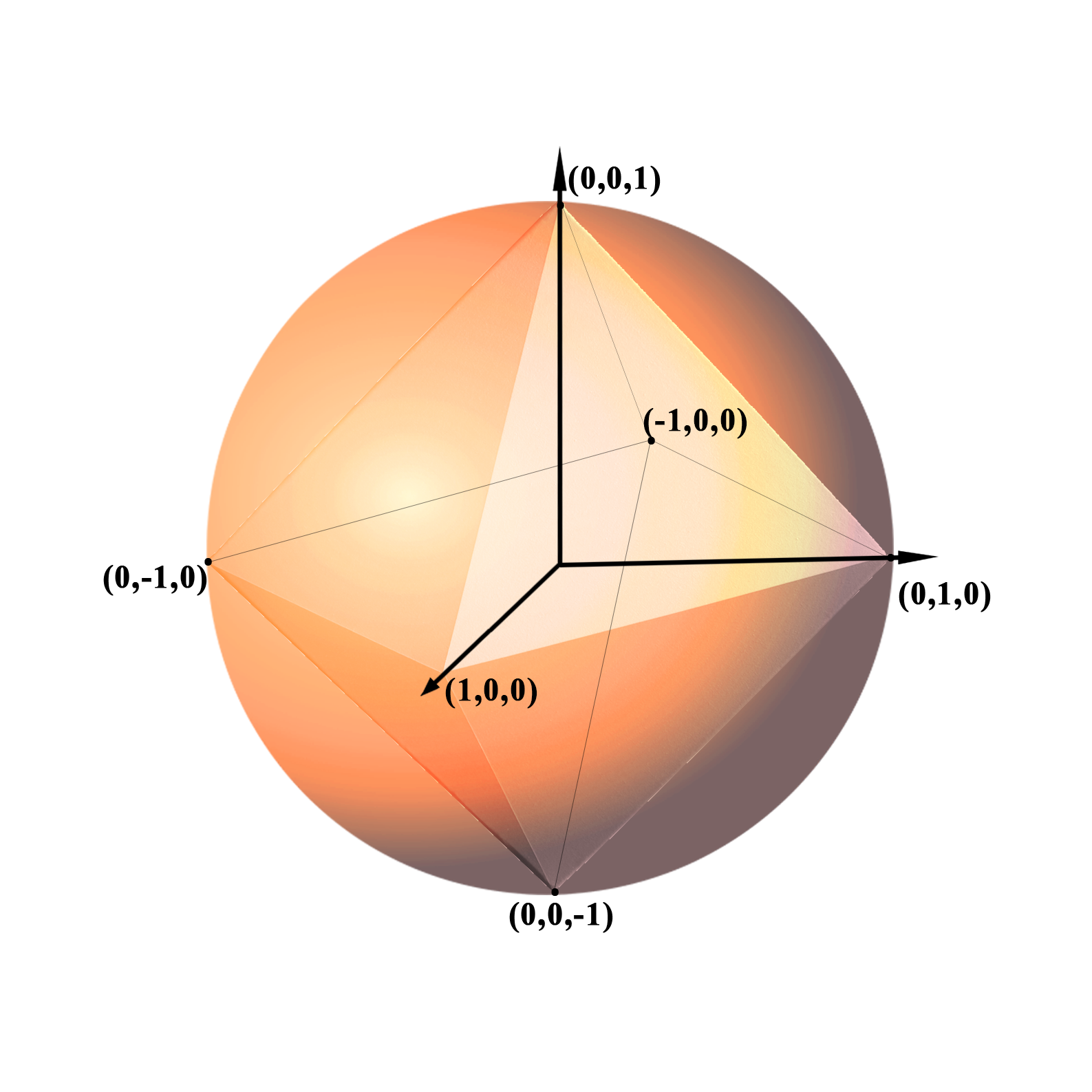}
\caption{\small{Within the Bloch ball $\mathcal B^2(\vec \theta)$, the octahedron $|\theta_1|+|\theta_2|+|\theta_3| \leq 1$ with vertices on the surface contains states for which the QPD corresponds to a joint probability distribution on measurement outcomes.}}
\label{pic}
\end{figure}

The eigenvalues of each of the three Pauli operators are $\pm 1$. Accordingly, 
consider three classical random variables $X_1,X_2,X_3$ each of which assumes values in $\mathcal X=\{1,-1\}$. 
Using the Margenau-Hill symmetrisation rule on Pauli operators define
\begin{equation}
\label{charac}
\phi (t_{1}, t_{2}, t_{3})= \frac{1}{3!} \Tr[\rho (\vec \theta)(\beta_{123}+\beta_{132}
+\beta_{213} +\beta_{231} +\beta_{312} +\beta_{321})],
\end{equation} 
 where $\beta_{abc}= e^{it_{a}\sigma_{a}}e^{it_{b}\sigma_{b}}e^{it_{c}\sigma_{c}}$ with $a,b,c=\{1,2,3\}$. Further simplification using 
 \begin{align}
 \label{identity}
 e^{i\sigma_{k}t_{k}}&= I_{2} \cos t_{k} + i\sigma_k \sin t_{k}, \thickspace \sigma^{2}_{k}= \mathbb{I}, \quad k= 1, 2, 3;\\
 \{\sigma_a, \sigma_b \}&= \delta_{ab} \sigma_c, \ [\sigma_a, \sigma_b]= 2i \epsilon_{abc} \sigma_c, \nonumber
 \end{align}
 implies that $\phi$ can written as
\begin{equation*}
\phi (t_{1}, t_{2}, t_{3})=\cos t_{1} \cos t_{2} \cos t_{3} + i\theta_{1} \sin t_{1} \cos t_{2} \cos t_{3} + \\ i\theta_{2} \cos t_{1} \sin t_{2} \cos t_{3} + i\theta_{3} \cos t_{1} \cos t_{2} \sin t_{3}.
\end{equation*}
In \cite{Sirsi07} $\phi$ was inverted to obtain the QPD
 \[
 p(x_1,x_2,x_3)=\frac{1}{8}(1+x_{1}\theta_1+x_{2}\theta_2+x_{3}\theta_3), \quad x_i \in \mathcal X, i=1,2,3.
 \]
 The QPD is non-negative only for those states $\vec \theta=(\theta_1,\theta_2,\theta_3)^T$ that, in addition to belonging to the Bloch ball $\mathcal B^2(\vec \theta)$, satisfy the inequality $|\theta_1|+|\theta_2|+|\theta_3| \leq 1$. The inequality characterises a octahedron in $\mathbb R^3$ with centre at $(0,0,0)$ within the Bloch ball with six vertices $(\pm 1, 0, 0)$, $(0, \pm 1, 0)$, $(0, 0, \pm 1)$ on the surface of the ball. Figure \ref{pic} provides a graphical representation.
 
 Thus for every state $\vec \theta=(\theta_1,\theta_2,\theta_3)^T$ inside the octahedron the function $p(x_1,x_2,x_3)$ is the joint distribution of classical random variables $(X_1,X_2,X_3)^T$, and we can thus \emph{prescribe} and accordingly interpret $p(x_1,x_2,x_3)$ as joint probabilities of outcomes of non-commuting Pauli measurement operators. 

\section{Pauli-like complete set of commuting operators for spin-1 system}
\label{sec:csco}
A spin-1 density matrix is characterized by eight independent parameters. Extending the methodology used for spin-1/2 system requires the representation of density matrix in a matrix basis which mimics the role played by Pauli-like operators, whose eigenstates form a complete set of MUBs. For spin-1 and higher-level spin systems it can be shown that using arbitrary $\mathfrak{su}(n)$ Lie algebra generators 
do not necessarily lead to a polytope similar to the spin-1/2 case.
Moreover, their physical implementation is not straightforward. We instead consider the MUB-driven operators $\mathcal A=\{\hat \alpha_j, j=1,\ldots,8\}$ proposed in \cite{SRao19}: 
	
{\footnotesize
	\begin{align*}
	\hat{{\alpha}}_{1}&=  {\sqrt{\frac{3}{2}}} \left(\begin{array}{ccc}
	1 & 0 & 0\\
	0 & 0 & 0\\
	0 & 0 & -1\\
	\end{array} \right),
	\qquad
	\hat{\alpha}_{2}=  \frac{1}{\sqrt 2} \left(\begin{array}{ccc} 
	1 & 0 & 0\\
	0 & -2 & 0\\
	0 & 0 & 1\\
	\end{array} \right),\\
	\hat{\alpha}_{3}&= {\frac{1}{\sqrt{2}}} \left(\begin{array}{ccc} 
	0 & -i\omega & i\omega^{2}\\
	i\omega^{2} & 0 & -i\omega\\
	-i\omega & i\omega^{2} & 0\\
	\end{array} \right), 
	\quad
	\hat{\alpha}_{4}= \frac{1}{\sqrt 2}\left(\begin{array}{ccc} 
	0 & -\omega & -\omega^{2}\\
	-\omega^{2} & 0 & -\omega\\
	-\omega & -\omega^{2} & 0\\
	\end{array} \right),\\ 
	\hat{\alpha}_{5}&={\frac{1}{\sqrt{2}}} \left(\begin{array}{ccc} 
	0 & -i  & i\omega^{2} \\
	i  & 0 & -i  \omega^{2} \\
	-i \omega & i \omega & 0\\
	\end{array} \right), 
	\quad
	\hat{\alpha}_{6}=\frac{1}{\sqrt 2} \left(\begin{array}{ccc} 
	0 & -1 & -\omega^{2}\\
	-1 & 0 & -\omega^{2} \\
	-\omega & -\omega & 0\\
	\end{array} \right),\\
	\hat{\alpha}_{7}& ={\frac{1}{\sqrt{2}}} \left(\begin{array}{ccc} 
	0 & -i \omega^{2} & i\omega^{2} \\
	i \omega & 0 & -i\\
	-i \omega & i  & 0\\
	\end{array} \right),
	\quad
	\hat{\alpha}_{8}= {\frac{1}{\sqrt{2}}} \left(\begin{array}{ccc} 
	0 & -\omega^{2} & -\omega^{2}\\
	-\omega & 0 & -1\\
	-\omega & -1 & 0\\
	\end{array} \right).
	\end{align*}
}
where $\omega= e^{2\pi i/3}$. Since $\Tr(\hat{\alpha}_{i} \hat{\alpha}_{j})= 3 \delta_{ij}$, $\mathcal A$ is an orthonormal operator basis for Hermitian matrices; it is a mutually disjoint, maximally commuting set since it can be partitioned into four sets $\mathcal A_1=\{\hat \alpha_1,\hat \alpha_2\}, \mathcal A_2=\{\hat \alpha_3,\hat \alpha_4\},\mathcal A_3=\{\hat \alpha_5,\hat \alpha_6\}, \mathcal A_4=\{\hat \alpha_7,\hat \alpha_8\}$ such that the operators within each $\mathcal A_i$ commute:$(\hat \alpha_j, \hat \alpha_{j+1})$ commute for $j=1,3,5,7$. This is the reason for them being referred to as a complete set of commuting operators (CSCOs). They are Pauli-like in the sense that their eigenbases are MUBs; we refer to \cite{SRao19} for details. The density matrix can be expressed as 
\[\rho(\vec \theta)= \frac{1}{3}[\mathbb{I}_{3} + \sum_{j= 1}^{8} \theta_{j} \hat{\alpha}_{j}], \quad \theta_{j}= \Tr[\rho(\vec \theta) \hat{\alpha}_{j}], \thinspace j=1,\ldots,8
.\]
The condition $\Tr [\rho(\vec \theta)^2] \leq 1$ implies that $\sum_{j=1}^8 \theta^2_j \leq 2$, and a similar seven-dimensional Bloch ball $\mathcal B^7(\vec \theta)$ of radius $\sqrt{2}$ in eight dimensions emerges. Bounds on the parameters are given by $-\sqrt{\frac{3}{2}} \leq \theta_{j} \leq \sqrt{\frac{3}{2}} $ when $j= 1, 3, 5, 7$, and $-{\sqrt{2}} \leq \theta_{j} \leq \frac{1}{\sqrt{2}}$ when  $j= 2, 4, 6, 8$. Since $\{\hat \alpha_j\}$ comprises of four sets of two commuting operators, we have three pairs of eigenvalues as measurement outcomes
shared between the operators, denoted as tuples $z_i=(x_i,y_i), i=1, 2, 3, 4$, where 
 \[ z_i \in \mathcal{Z}=
\left\{\left(\sqrt{\frac{3}{2}}, \sqrt{\frac{1}{2}}\right), \left(0, -\frac{2}{\sqrt{2}}\right),
	\left(-\sqrt{\frac{3}{2}}, \sqrt{\frac{1}{2}}\right)\right\}.
	\]
	

\section{Quasiprobabilities for spin-1 system}
\label{sec:spin1}
We consider four measurement operators $M_i, i=1,2,3,4$, instead of $n^2-1=8$, where each $M_i$ is defined using the two commuting operators in $\mathcal A_i$. To our knowledge, such a construction is not possible with other operator basis (for e.g. Gell-Mann matrices). For a fixed $\vec t =(t_1,\ldots,t_8)$, let 
\begin{align*}
M_1&= \hat{\alpha}_{1} t_{1}+\hat{\alpha}_{2}t_{2},\quad  M_2= \hat{\alpha}_{3} t_{3}+\hat{\alpha}_{4}t_{4},\\
M_3&= \hat{\alpha}_{5} t_{5}+\hat{\alpha}_{6}t_{6}, \quad M_4= \hat{\alpha}_{7} t_{7}+\hat{\alpha}_{8}t_{8}. 
\end{align*}
Unlike the case for spin-1/2 systems it is not straightforward to explicitly compute the expression the characteristic function using the Margenau-Hill symmetrization rule (or any rule for that matter) since a simplifying relation such as \eqref{identity} is unavailable. 

Another advantage in using the CSCOs is that we can analyse the characteristic function for the eight-dimensional QPD in a modular manner:
characteristic functions using pairs $M_i, M_j$ can first be analysed and then for triples $M_i,M_j,M_k$, which then enable a straightforward derivation of the QPD. Relatedly, we can thus consider 4 bivariate random variables $\vec X=((X_1, Y_1)^T,(X_2, Y_2)^T, (X_3, Y_3)^T, (X_4, Y_4)^T)^T$, as opposed to a single 8-dimensional random vector. Each bivariate random variable $(X_i, Y_i)^T$ assumes values  $z_i=(x_i,y_i) \in \mathcal{Z}$. 

The operator $M_1$ is diagonal and Hermitian. Since the transformation from $M_i$ to $M_j$ corresponds to a unitary transformation from one MUB basis to another, we see that $M_2, M_{3}$ and $M_{4}$ can also be diagonalised, respectively, with unitary transformations 

 {\footnotesize
\begin{align*}
 U_2=\frac{1}{\sqrt{3}}\left(
\begin{array}{ccc}
  1 & 1 & 1  \\
1 & \omega & \omega^{2} \\
1 & \omega^{2} & \omega \\
\end{array}\right), \thinspace
U_3 =\frac{1}{\sqrt{3}}	 \left(
\begin{array}{ccc}
  1 & \omega^{2} & 1  \\
1 & 1 & \omega^{2} \\
1 & \omega & \omega \\
\end{array}\right),\thinspace
U_4 =\frac{1}{\sqrt{3}}\left(
\begin{array}{ccc}
  1 & \omega & 1  \\
1 & \omega^{2} & \omega^{2} \\
1 & 1 & \omega \\
\end{array}\right).
\end{align*}
}
For benefit of exposition we describe the construction in an incremental fashion: aided by the decomposition of $\mathcal A$ into subsets containing commuting operators and availability of explicit unitary transformations between MUB bases, we detail how the four-dimensional marginal QPDs (corresponding to any two $M_i$ and $M_j$) of the eight-dimensional QPD we seek can be defined; the methodology extends to six-dimensional marginal QPDs obtained using three operators. Consider the characteristic function
\begin{equation*}
    \phi(t_1, t_2, t_3, t_4)= \frac{1}{2} \Tr[\rho(\vec \theta)(e^{iM_{1}}e^{iM_{2}}+e^{iM_{2}}e^{iM_{1}})]
    \end{equation*}
    defined only using $M_1$ and $M_2$. Since $\hat{\alpha}_{3}= U^{\dagger}_{2} \hat{\alpha}_{1} U_2$ and $\hat{\alpha}_{4}= U^{\dagger}_{2} \hat{\alpha}_{2} U_2$, we have
    \begin{equation}
    \label{eq:MUB}
    \phi(t_1, t_2, t_3, t_4)= \frac{1}{2} \Tr[\rho(\vec \theta)(e^{iM_{1}}U^{\dagger}_{2} e^{iM^{d}_{2}}U_2+U^{\dagger}_{2}e^{iM^{d}_{2}}U_{2} e^{iM_{1}})],
    \end{equation}
    where 
    $M^{d}_{2}= \hat{\alpha}_{1} t_{3}+\hat{\alpha}_{2}t_{4}$.
    Denote the joint eigenbases of $\hat \alpha_1$ and $\hat \alpha_2$ (since they commute) in $\mathcal A_1$ as $|z_{1} \rangle$, where $z_1=(x_1,y_1) \in \mathcal Z$ is the corresponding eigenvalue pair. That is, $z_1$ can assume three values in $\mathcal Z$ and $|z_{1} \rangle$ thus generically denotes any of the three eigenvectors that are common to both $\hat \alpha_1$ and $\hat \alpha_2$. Similarly consider $|z_{2} \rangle$ from $\mathcal A_2$. 
	    Following the technique used in \cite{Devi1994}, we have
	 \begin{align*}  
	  \phi(t_1, t_2, t_3, t_4)= \sum_{z_{1}, z_{2}} e^{i(x_1t_1+y_1t_2+x_2t_3+y_2t_4)} \text{Re}\{\langle z_2|U_{2} |z_1\rangle \langle z_1|\rho(\vec \theta)U^\dagger_{2} |z_2\rangle\},
	     \end{align*}
	     where $\text Re\{y\}$ denotes the real part of the complex number $y$. Observing the right hand side of $\phi$, we can thus, without explicitly inverting the characteristic function, prescribe the four-dimensional QPD
	     \begin{align*}
	     p(z_{1}, z_{2})&= \text{Re}\{\langle z_2|U_{2} |z_1\rangle \langle z_1|\rho(\vec \theta)U^\dagger_{2} |z_2\rangle\}\\
	     &=\frac{1}{9}\left[1+x_1 {\theta_{1}} + y_1{\theta_{2}}+ x_2{\theta_{3}} + y_2 {\theta_{4}}\right]
	    \end{align*}
following some algebra. The QPD above is the four-dimensional marginal of the eight-dimensional QPD of interest. Using the unitary operators $U_j, j=2,3,4$, the other four-dimensional QPDs $p(z_k,z_j),j,k=2,3,4$ with $j \neq k$ can then be obtained as
$$p(z_{j}, z_{k})= \text{Re}\{\langle z_k|U_{k}U^\dagger_j |z_j\rangle \langle z_j|U_j\rho(\vec \theta)U^\dagger_{k} |z_k\rangle\}.$$ 
The above program can be carried out for operators in $\mathcal A_1,\mathcal A_2,\mathcal A_3$ in almost identical fashion using operators $M_1,M_2,M_3$ to obtain the six-dimensional marginal QPD
	 \begin{align*}
	     p(z_{1},z_{2},z_{3})= \frac{1}{3^3}& \left[1 + x_1 {\theta_{1}} + y_1{\theta_{2}}+ x_2{\theta_{3}} + y_2 {\theta_{4}} + x_3 {\theta_{5}} + y_3{\theta_{6}} \right],
\end{align*}
using which the other six-dimensional marginals QPD can be derived using the corresponding unitary operators. Finally, using all four measurement operators $M_i=1,2,3,4$, the characteristic function for the eight-dimensional QPD can written as
\begin{equation*}
    \phi(\vec t)= \frac{1}{4!}\sum _{\pi \in \Pi_4} \Tr\left[\rho(\vec \theta) \left(\beta(\pi(1)\pi(2)\pi(3)\pi(4))\right)\right],
\end{equation*}
where $\beta(abcd)= e^{i\hat{M}_{a}}e^{i\hat{M}_{b}}e^{i\hat{M}_{c}}e^{i\hat{M}_{d}}$ and $a,b,c,d \in \{1,2,3,4\}$. Following the above steps with unitaries $U_2,U_3,U_4$ enables the definition of the required eight-dimensional QPD
\begin{equation*}
    \phi(\vec t)= \sum _{z_1, z_2, z_3, z_4 \in \mathcal Z} e^{i(x_1t_1+y_1t_2+x_2t_3+y_2t_4+x_3t_5+y_3t_6+x_4t_7+y_4t_8)} p(z_1, z_2, z_3, z_4),
\end{equation*}
with
\begin{align*}
{\label{MH2}}
p(z_1,z_2,z_3,z_4)= \frac{1}{3^4}& \left[1 + x_1 {\theta_{1}} + y_1{\theta_{2}}+ x_2{\theta_{3}} + y_2 {\theta_{4}} + x_3 {\theta_{5}} + y_3{\theta_{6}}+ x_4{\theta_{7}} + y_4{\theta_{8}} \right].
\end{align*}
The form of the QPD for a spin-1 system is similar to that for spin-1/2 case. This is a consequence of employing Pauli-like CSCOs. 
\subsection{Geometric description}
\label{sec:spin1geom}
The set of states within the Bloch ball $\mathcal B^7(\vec \theta)$ for which the QPD is a joint distribution on eight classical random variables is characterised by the linear inequality
$$x_1 {\theta_{1}} + y_1{\theta_{2}}+ x_2{\theta_{3}} + y_2 {\theta_{4}} + x_3 {\theta_{5}} + y_3{\theta_{6}}+ x_4{\theta_{7}} + y_4{\theta_{8}}  \geq -1.$$ 
The set is convex polytope in $\mathbb R^8$. The use of $4 \times 3=12$ MUB basis vectors used to construct the operators $M_1, M_2,M_3$  leads to the polytope to have 12 vertices on the surface of $\mathcal B^7(\vec \theta)$ that are the only pure states given by $1/\sqrt{2}$ times the coordinates
\begin{align*}
&(\sqrt{3},1,\vec 0_6), 
(-\sqrt{3},1,\vec 0_6), (\vec 0_6, \sqrt{3},1),(\vec 0_4,-\sqrt{3},1,\vec 0_2), \nonumber \\
& (\vec 0_6, -\sqrt{3},1),(0,-2,\vec 0_6), (\vec 0_3,-2,\vec 0_4), (\vec 0_5,-2,\vec 0_2), \nonumber \\
& (\vec 0_7,-2), (\vec 0_2,\sqrt{3},1,\vec 0_4), (\vec 0_2,-\sqrt{3},1,\vec 0_4), (\vec 0_4,\sqrt{3},1,\vec 0_2),
\end{align*}
where $\vec 0_r$ denotes a vector of $r$ zeroes. We observe that each vertex has two vertices with which it subtends an angle $2 \pi/3$ at the origin, and is orthogonal to the rest; the two vertices are the `diametrically opposite' points on the Bloch sphere and are linked by an $SU(3)$ rotation. The three vertices comprise an equilateral triangle; consider for example, the triplet $(\sqrt{3},1,\vec 0_6), 
(-\sqrt{3},1,\vec 0_6), (0,-2,\vec 0_6)$. Consequently, we note that each vertex is formed by four mutually orthogonal equilateral
triangular planes in the Bloch sphere, which implies that the polytope has $3^4=81$ faces. \\ \\
There are no edges between vertices that correspond to vectors from the same basis set, and edges of equal length are formed with every vertex outside the basis set. Recall that the MUB comprises 4 basis sets containing 3 vectors each. Thus a vertex from the first MUB basis can share an edge with $3^2$ vertices that are orthogonal to it and do not belong to the same MUB basis. This is true for each vertex in the first MUB set and there are hence $3^3$ edges involving vertices from the first MUB set. In similar fashion, there are $3^2(3-1)=3^2 \times 2$ edges involving vertices from the second MUB set, and so on. The total number of edges of the polytope is hence $3^3(3+1)/2=54$. \\ \\
As a geometric object, the convex polytope matches the one described in \cite{Galvao04} using the discrete Wigner QPD and in \cite{Beng05} using the so-called probability coordinates $\vec{p}$. They are related in the following manner: elements of the Bloch vector $\vec \theta$ that coordinatises the Bloch ball are expectation values $\theta_i=\Tr(\rho (\vec \theta)\hat \alpha_i)$ of CSCOs $\hat \alpha_i, i=1,\ldots,8$; if instead projection operators $\Pi_1,\ldots,\Pi_8$ (along with the identity operator) corresponding to the MUB basis vectors are used, their expectation values $p_i=\Tr(\rho (\vec \theta) \Pi_i)$ constitute the probability coordinates $\vec p$ with $\sum_{i=1}^8 p_i +(1-\sum_{i=1}^8 p_i)=1$. In effect, this amount to a specific reparameterisation of the map $\theta \mapsto \rho(\theta)$, whose image is thus preserved. The projectors $\Pi_i$, however, do not form an orthonormal set and it is not possible to provide Bloch vector-like coordinates to a density matrix; moreover, it is not straightforward to physically realise projection operators. In contrast, the CSCOs comprise an orthonormal operator basis with which the density matrix is provided interpretable coordinates, and, as will be seen shortly, can be physically realised. 

{\subsection{Physical realization of measurement operators}} The Hamiltonian associated with the first MUB of spin-1 system is a linear combination of $\hat{\alpha}_{1}$ and $\hat{\alpha}_{2}$. For spin systems, it is natural to consider  the irreducible spherical tensor operator basis $\tau^k_q$ rank $k$ in the $2j+1$ dimensional spin space with projection $q$ along the axis of quantization in the real 3-dimensional space. The matrix elements of $\tau^{k}_{q}$ are 
\begin{equation*}
\langle jm' |\tau^{k}_{q}(\vec{J}) | jm \rangle= \sqrt{2k+1} C(jkj; mqm'),
\end{equation*} 
where $C(jkj;mqm')$ are the Clebsch--Gordan coefficients. $\tau^{k}_{q}$s satisfy
\begin{equation*}
	\Tr({\tau^{k^{\dagger}}_{q}\tau^{k^{'}}_{q^{'}}})= (2j+1)\,\delta_{kk^{'}} \delta_{qq^{'}}, \quad \tau^{k^{\dagger}}_{q} = (-1)^{q}\tau^{k}_{-q},
\end{equation*}
where the normalization has been chosen so as to be in agreement with Madison convention. 
Then, $\hat{\alpha}_{1}= \tau^{1}_{0}= \sqrt{\frac{3}{2}} J_{z}$, $\hat{\alpha}_{2}= \tau^{2}_{0}= \frac{3J^{2}_{z}-J^{2}}{\sqrt{2}}$. The expectation values of $\hat{\alpha}_{1}$ and $\hat{\alpha}_{2}$ are respectively associated with the first and second order moments of $J_{z}$ and constitute experimentally measurable parameters. The Hamiltonian associated with second MUB is obtained from the Fourier transformation of the first basis. Similarly, transition from second to third MUB is obtained from one-axis twisting $e^{-iS^{2}_{z}t}$ for $t=2\pi/3$ and from second to fourth MUB for $t=4\pi/3$ \cite{kit}. Thus, the complete state determination results in determining the parameters $\theta_i$, $i= 1, \ldots 8$, which is optimally done using the complete set of commuting operators $\hat{\alpha}_{i}$.\\
Experimentally this corresponds to the application of external electric quadrupole field in addition to the dipole magnetic field in the Stern-Gerlach setup. The Hamiltonian is diagonal in the first MUB has the form
${\mathcal{H}}_1= h_{0} \mathbb{I}+h_1 {\hat{\alpha}}_{1}+h_2 {\hat{\alpha}}_{2}$; alternatively, in terms of spherical tensors, $\mathcal{H}_1= \sum_{k}^{2} h^{k}_{0} {\tau^{k}}^{\dagger}_{0}.$
 What one experimentally measures is the expectation value of the Hamiltonian $$\Tr(\rho \mathcal{H}_1)= h_0+h_1 \theta_1+h_2 \theta_2.$$ Unitary transformations connecting different MUBs from the canonical basis is parametrized by a single parameter $\phi$, ${\hat{U}}_{i}=e^{i {\mathcal{H}_{i}} \phi_i}$, where ${\mathcal{H}}_i$ is the Hamiltonian diagonal in the $i^{th}$ basis and $\phi_i= 2\pi/3, 4\pi/3, 2\pi$   for $i=2, 3, 4$ respectively.  
\section{Quasiprobabilities for spin-3/2 system}
\label{sec:spin3/2}
For spin-3/2 systems of dimension $n=4$, and $n^2-1=15$, we briefly describe the construction of a QPD and the ensuing geometric picture of states along the lines of what is done for spin-1 systems. Following the method proposed in \cite{SRao19}, the CSCOs for spin-3/2 system is explicitly given by
{\footnotesize
\begin{align*}
{\hat{\beta}}_{1} =		 \frac{1}{\sqrt{5}}\left(
\begin{array}{cccc}
  3 & 0 & 0 & 0 \\
0 & 1 & 0 & 0\\
0 & 0 & -1 & 0\\
0 & 0 & 0 & -3\\
\end{array}\right), 
\quad
{\hat{\beta}}_{2} =		 \left(
\begin{array}{cccc}
  1 & 0 & 0 &0  \\
0 & -1 & 0& 0 \\
0& 0& -1 &0\\
0& 0& 0 & 1\\
\end{array}\right), 
\quad
{\hat{\beta}}_{3} =		 \frac{1}{\sqrt{5}}  \left(
\begin{array}{cccc}
  1 & 0 & 0 & 0\\
0 & -3 & 0 & 0 \\
0 & 0 & 3 &0 \\
0 & 0 & 0 & -1\\
\end{array}\right), 
\end{align*}
}
{\footnotesize
\begin{align*}
{\hat{\beta}}_{4} =		 \frac{1}{\sqrt{5}}\left(
\begin{array}{cccc}
  0 & 1 & 2 & 0 \\
1 & 0 & 0 & 2\\
2 & 0 & 0 & 1\\
0 & 2 & 1 & 0\\
\end{array}\right), 
\quad
{\hat{\beta}}_{5} =		 \left(
\begin{array}{cccc}
  0 & 0 & 0 &1  \\
0 & 0 & 1& 0 \\
0& 1& 0 &0\\
1& 0& 0 & 0\\
\end{array}\right), 
\quad
{\hat{\beta}}_{6} =		 \frac{1}{\sqrt{5}}  \left(
\begin{array}{cccc}
  0 & 2 & -1 & 0\\
2 & 0 & 0 & -1 \\
-1 & 0 & 0 &2 \\
0 & -1 & 2 & 0\\
\end{array}\right), 
\end{align*}
}
{\footnotesize
\begin{align*}
{\hat{\beta}}_{7} =		 \frac{1}{\sqrt{5}}\left(
\begin{array}{cccc}
  0 & -i & -2i & 0 \\
i & 0 & 0 & -2i\\
2i & 0 & 0 & -i\\
0 & 2i & i & 0\\
\end{array}\right), 
\quad
{\hat{\beta}}_{8} =		 \left(
\begin{array}{cccc}
  0 & 0 & 0 &-1  \\
0 & 0 & 1& 0 \\
0& 1& 0 &0\\
-1& 0& 0 & 0\\
\end{array}\right), 
\quad
{\hat{\beta}}_{9} =		 \frac{1}{\sqrt{5}}  \left(
\begin{array}{cccc}
  0 & -2i & i & 0\\
2i & 0 & 0 & i \\
-i & 0 & 0 &-2i \\
0 & -i & 2i & 0\\
\end{array}\right), 
\end{align*}
}
{\footnotesize
\begin{align*}
{\hat{\beta}}_{10} =		 \frac{1}{\sqrt{5}}\left(
\begin{array}{cccc}
  0 & -i & 2 & 0 \\
i & 0 & 0 & -2\\
2 & 0 & 0 & i\\
0 & -2 & -i & 0\\
\end{array}\right), 
\quad
{\hat{\beta}}_{11} =		 \left(
\begin{array}{cccc}
  0 & 0 & 0 &i  \\
0 & 0 & i& 0 \\
0& -i& 0 &0\\
-i& 0& 0 & 0\\
\end{array}\right), 
\quad
{\hat{\beta}}_{12} =		 \frac{1}{\sqrt{5}}  \left(
\begin{array}{cccc}
  0 & -2i & -1 & 0\\
2i & 0 & 0 & 1 \\
-1 & 0 & 0 &2i \\
0 & 1 & -2i & 0\\
\end{array}\right), 
\end{align*}
}
{\footnotesize
\begin{align*}
{\hat{\beta}}_{13} =		 \frac{1}{\sqrt{5}}\left(
\begin{array}{cccc}
  0 & 1 & -2i & 0 \\
1 & 0 & 0 & 2i\\
2i & 0 & 0 & -1\\
0 & -2i & -1 & 0\\
\end{array}\right), 
\quad
{\hat{\beta}}_{14} =		 \left(
\begin{array}{cccc}
  0 & 0 & 0 &i  \\
0 & 0 & -i& 0 \\
0& i& 0 &0\\
-i& 0& 0 & 0\\
\end{array}\right), 
\quad
{\hat{\beta}}_{15} =		 \frac{1}{\sqrt{5}}  \left(
\begin{array}{cccc}
  0 & 2 & i & 0\\
2 & 0 & 0 & -i \\
-i & 0 & 0 & -2 \\
0 & i & -2 & 0\\
\end{array}\right), 
\end{align*}
}
where $\Tr(\hat{\beta}_{i}\hat{\beta}_{j})= 4\delta_{ij}$. The density matrix can be expressed as
\[
\rho(\vec \theta)= \frac{1}{4}[\mathbb{I}_{4} + \sum_{j= 1}^{15} \theta_{j} \hat{\beta}_{j}],
\] where $\theta_{j}= \Tr[\rho(\vec \theta) \hat{\beta}_{j}]$. 
Denote by $\mathcal A=\cup_{i=1}^5\mathcal A_i$ where $\mathcal A_1=\{\hat \beta_1,\hat \beta_2,\hat \beta_3\}, \mathcal A_2=\{\hat \beta_4,\hat \beta_5,\hat \beta_6\}, \mathcal A_3=\{\hat \beta_7,\hat \beta_8,\hat \beta_9\}, \mathcal A_4=\{\hat \beta_{10},\hat \beta_{11},\hat \beta_{12}\}, \mathcal A_5=\{\hat \beta_{13},\hat \beta_{14},\hat \beta_{15}\}$; we note that there are thus five sets $\mathcal A_i$ each consisting of three commuting operators. Here, four eigenvalue triples are shared between the operators, $z_i=(w_i, x_i, y_i)$, $i= 1, 2, 3, 4, 5$ and are given by, 
 \[ z_i \in \mathcal{Z}=
\left\{
	\left(\sqrt{\frac{9}{5}}, 1, \frac{1}{\sqrt{5}}\right), \left(\frac{1}{\sqrt{5}}, -1, -\sqrt{\frac{9}{5}}\right), \left(-\frac{1}{\sqrt{5}}, -1, \sqrt{\frac{9}{5}}\right), \left(-\sqrt{\frac{9}{5}}, 1, -\frac{1}{\sqrt{5}}\right)\right\}.
	\]
For a fixed $\vec{t}=(t_1,\ldots,t_{15})$, define the measurement operators
\begin{align*}
M_{1}&= \hat{\beta}_{1} t_{1}+\hat{\beta}_{2}t_{2}+\hat{\beta}_{3}t_{3},\quad  M_2= \hat{\beta}_{4} t_{4}+\hat{\beta}_{5}t_{5}+\hat{\beta}_{6}t_{6},\quad  M_3= \hat{\beta}_{7} t_{7}+\hat{\beta}_{8}t_{8}+\hat{\beta}_{9}t_{9},\\
M_4&= \hat{\beta}_{10} t_{10}+\hat{\beta}_{11}t_{11}+\hat{\beta}_{12} t_{12}, \quad M_5= \hat{\beta}_{13}t_{13}+\hat{\beta}_{14}t_{14}+\hat{\beta}_{15} t_{15}.
\end{align*}
The unitary transformations $\hat{U}_{j}, j=2, 3, 4, 5$ take $M_{2}$, $M_{3}$, $M_{4}$, $M_{5}$ to their diagonal form are known \cite{zei}.
Along the lines of what has was done for spin-1 system, the resulting characteristic function for spin-3/2 system is 
\begin{equation*}
    \phi(\vec t)= \frac{1}{5!}\sum _{\pi \in \Pi_4} \Tr\left[\rho(\vec \theta) \left(\zeta(\pi(1)\pi(2)\pi(3)\pi(4)\pi(5))\right)\right],
\end{equation*}
where $\zeta(abcde)= e^{iM_{a}}e^{iM_{b}}e^{iM_{c}}e^{iM_{d}}e^{iM_{e}}$ and $a,b,c,d,e \in \{1,2,3,4,5\}$. 
Let $\vec \theta=(\theta_1,\ldots,\theta_{15})^T$ with $\vec \theta=(\vec \theta_1,\vec \theta_2,\ldots,\vec \theta_5)^T$ where $\vec \theta_1=(\theta_1,\theta_2,\theta_3)$, and so on. Recall that the triples $z_i=(w_i,x_i,y_i), i=1,\ldots,5$ can be viewed as three-dimensional vectors. Following the steps laid out for spin-1 results in the fifteen-dimensional QPD
\[
p(z_1, z_2, z_3, z_4, z_5) =\frac{1}{1024}\left[1+z_1\cdot \vec \theta_1+\cdots+z_5\cdot \vec \theta_5\right],\quad z_i \in \mathcal Z,
\]
where $a \cdot b$ is the dot product between vectors $a$ and $b$. Each measurement operator $M_i, i=1,\ldots,5$ will engender a three-dimensional marginal of the joint QPD. 

The convex polytope within which the QPD is a valid joint distribution on classical random variables is given by the inequality $z_1\cdot \vec \theta_1+\cdots+z_5\cdot \vec \theta_5 \geq -1$. Along the line of reasoning used for the spin-1 case, we note that the polytope has $5 \times 4=20$ vertices; each vertex is equidistant from the 3 other vertices within the MUB set, but does not share an edge with any of them; it shares an edge, and is orthogonal, with the rest of the  vertices. The polytope thus has $4^5=1024$ faces, and $4^3(4+1)/2=160$ edges. 
\section{Higher-order spin systems}
\label{sec:spinj}
Method of construction of the QPD described can be used for higher-order spin-$j$ systems if the corresponding set of CSCOs are available. The CSCOs, in principle, can be constructed since a complete set of MUBs is known to exist when the dimension $n=2j+1$ is a prime or power of a prime. For such $n$, we can consider the irreducible tensor operators $\tau^{k}_{q}$ discussed above, with $\tau^{0}_{0}= \mathbb{I}_n$ being the identity operator. The matrix elements of diagonal operators are $\langle jm'|\tau^{k}_{0}|jm \rangle= \sqrt{2k+1}  C(jkj;m0m') $. 

Thus in the canonical basis of Hilbert space of dimension $n=2j+1$, we know the $2j$ diagonal matrices. It is possible to generate CSCOs from the unitary transformations connecting different MUB sets\cite{SRao19}, resulting in $n+1$ sets $\mathcal A_i, i=1,\ldots,n+1$ of operators, where each $\mathcal A_i$ contains $n-1$ commuting operators. The $n+1$ measurement operators $M_i,i=1,\ldots,n+1$ are then constructed taking linear combinations of the operators within each $\mathcal A_i$. The CSCOs determine the Bloch vector $\vec \theta=(\theta_1,\ldots,\theta_{n^2-1})$ through the corresponding density matrix representation. Inspection of our method reveals that the requirement
to extend this to an arbitrary finite dimensional system is that a complete set of
MUBs is known to exist. For such systems, physical realization amounts to the identification of a
suitable Hamiltonian which plays a role similar to the
multipole fields used for spin-$j$ systems.

With $\vec v_i=(v_{1i},\ldots,v_{(n-1)i}), i=1,\ldots,n+1$ and a commensurate partitioning of the vector $\vec \theta$ as $\vec \theta=(\vec \theta_1,\ldots,\vec \theta_{n+1})^T$, the general form of the $(n^2-1)$-dimensional QPD for an $n$-dimensional system then is
\[
p(\vec v_1,\ldots,\vec v_{n+1})=\frac{1}{n^{n+1}}\left[1+\vec v_1 \cdot \vec \theta_1+\cdots+\vec v_{n+1} \cdot \vec \theta_{n+1}\right],
\]
where the $\vec v_i$ assume values in a set consisting of $n$ elements, where each element is an $(n-1)$-dimensional vector. The set of states within the Bloch ball $\mathcal B^{n^2-2}(\vec \theta)$ for which the QPD is non-negative is given by the inequality $\vec v_1 \cdot \vec \theta_1+\cdots+\vec v_{n+1} \cdot \vec \theta_{n+1} \geq -1$. The convex polytope has $n(n+1)$ vertices, $n^{n+1}$ faces, and has $n^3(n+1)/2$ edges.

\section{Acknowledgements}
HSS thanks the Department of Science and Technology (DST), India for the grant of INSPIRE Fellowship. KB acknowledges partial support from grants NSF DMS grants 1613054, 2015374 and NIH R37-CA214955.

\section*{References}
\bibliography{joint1}
\end{document}